# Phase coexistence and interrupted 1<sup>st</sup> order transition in magnetic shape memory alloys

P Chaddah*

RR Centre for Advanced Technology, Indore 452013, India.

*Current theoretical studies on structural and magnetic properties of functional Ni-Mn-Z (Z = Ga, In, Sn) Heusler alloys address the origin of the structural transition from the austenite to martensite, and also address the dominant contribution to the latent heat associated with this magneto-structural transition. This should help understand the origin of kinetic arrest of 1<sup>st</sup> order transitions.*

Phase coexistence down to the lowest temperatures, of two competing phases separated by a 1<sup>st</sup> order transition, was highlighted in the case of half-doped manganites [1,2]. Phase-coexistence in the manganites with colossal magneto-resistance (CMR) had been believed to be an inhomogeneous ground state. Similar behavior was shown in doped $CeFe_2$ alloys, and was attributed to the "kinetic arrest" of the 1<sup>st</sup> order ferromagnetic (FM) to anti-ferromagnetic (AFM) transition after it had progressed partially [3]. It was proposed that various other predictions of the "kinetic arrest" phenomenon should be seen in half-doped manganites, and this was confirmed [4]. Theorists have conceded "kinetic arrest" as a possible explanation for the behavior observed in manganites [5], but this was not investigated by detailed studies because the emphasis of theorists had by then shifted away from the CMR manganites.

This alternative explanation of phase-coexistence being a manifestation of a disorder-broadened first-order magnetic phase transition interrupted by the arrest of kinetics is gaining ground. This is because phase coexistence down to the lowest temperatures, due to broad 1<sup>st</sup> order magnetic transitions being interrupted by kinetic arrest, has been reported in many other materials. Some of these other families of materials, where this phenomenon has been established using various techniques including the specially designed protocol CHUF [4], are represented by $Gd_5Ge_4$ [6,7], doped $Mn_2Sb$ [8,9] doped FeRh [10], Ni-Mn based MSMAs [11,12], Ta-doped $HfFe_2$ [13], and Cobaltites [14,15]. Of these, the Ni-Mn based MSMAs have been the most extensively studied by many experimental groups because of their potential for applications. Also, amongst all the materials listed above the Ni-Mn based materials are now being investigated in some detail by theorists. It is envisaged that this theoretical effort could provide a testing ground of ideas that were proposed as we developed the phenomenology of a disorder-broadened first-order magnetic phase transition, interrupted by the arrest of kinetics. Specifically, disorder-broadening has been ascribed to the transition being driven by short range interactions, and kinetic arrest has been ascribed to the latent heat being weakly linked to the thermal conduction process.

I now mention these recent theoretical works on Ni-Mn-Z (Z = Ga, In, Sn) Heusler alloys, which are hopefully preludes to sustained ongoing efforts. There is a magneto-structural instability in these materials, and the austenite-martensite transition is accompanied by a sharp drop of magnetization attributed to a competition between ferromagnetic and antiferromagnetic interactions.

In a recent work by Comtesse et al [16], Ni-Mn-Z (Z = Ga, In, Sn) Heusler alloys are studied by first-principles and Monte Carlo methods, and the effect of substituting Co on Ni site is investigated. Tan et al [17] have earlier reported first-principles calculations on Ni-Mn-In as the ratio of Mn/In is varied. They report a critical value of 0.3 nm for Mn-Mn interatomic distance that corresponds to the crossover between ferromagnetic and antiferromagnetic exchange interactions between Mn atoms on the two sites. Tan et al [17] also conclude that the density of states at the Fermi level triggers the martensitic transformation; Comtesse et al [16] show that the martensite is stabilized by the creation of a pseudo-gap at $E_F$. While the itinerant electrons determine the martensitic transformation temperature, Comtesse et al [16] conclude that the interaction of the localized electronic orbitals determines the jump ΔM in magnetization, and thus the latent heat of the phase transition. We must note that both the magnetic and the structural transition are simultaneously interrupted in these materials [18]; the interaction driving the dynamics of the magneto-structural transition may need to be identified.

In a very recent paper, Wang et al [19] report first principles calculations on substituted Ni-Mn-Sn (the Ni-Z-Mn-Sn system with Z = Co, Fe, Mn and Cr substituted on Ni site) where the field-induced martensite to austenite transition can be observed. They use the Bethe-Slater model, with direct exchange between neighbouring atoms (and the overlap of neighbouring orbitals) causing the magnetic ordering, to study the magnetic transition. This supports the conclusion of Comtesse et al [16] that the interaction of the localized electronic orbitals determines the latent heat of the magnetic transition.

We now consider some of the conceptual issues that arise when a $1^{st}$ order magnetic transition is 'interrupted'. First, it has been argued [20] that a broad $1^{st}$ order transition will be observed when the interaction driving the transition is short range. Because only then can one ascribe transition temperatures $T_C$, or critical fields $H_C$, to localized spatial regions. Second, what conditions can cause the 'interruption'? This is a much more complex issue.

The concept of an 'interruption' is well accepted in the freezing transition of a liquid because the change in density involves motion of atoms over long distances, and 'interruption' (or glass-formation) occurs when the velocity is so low that the density cannot change over experimental time scales. We do not have a conceptual counterpart of this to suggest why a $1^{st}$ order magnetic transition is getting interrupted. Chaddah and Banerjee [21] had proposed an alternate criterion viz. a $1^{st}$ order transition will be interrupted if the heat conduction mechanism is weakly coupled to the latent heat. Since electronic heat conduction is due to itinerant electrons at the Fermi surface, this would require that the magnetic ordering is caused by localized orbital electrons; that the localized electronic orbitals determine the magnetic entropy change associated with the isothermal metamagnetic transition, as is concluded by Comtesse et al [16].

Clearly more detailed theoretical studies on the Ni-Z-Mn-Sn system with Z = Co, Fe, Mn and Cr substituted on Ni site, are eagerly awaited. Specifically, is the latent heat of the 1$^{st}$ order magneto-structural transition dominated by the magnetic part, and is it thus associated with orbital electrons? Is the transition really 'driven' by the electrons at $E_F$ [17]?

*Since retired; Email: chaddah.praveen@gmail.com